\documentstyle[twocolumn,aps,psfig]{revtex}

\begin{document}
\draft
\flushbottom
\twocolumn[
\hsize\textwidth\columnwidth\hsize\csname @twocolumnfalse\endcsname

\title{ Fractal extra dimension in Kaluza-Klein theory}
\author{Igor I. Smolyaninov }
\address{ Department of Electrical and Computer Engineering \\
University of Maryland, College Park,\\
MD 20742}
\date{\today}
\maketitle
\tightenlines
\widetext
\advance\leftskip by 57pt
\advance\rightskip by 57pt

\begin{abstract}
Kaluza-Klein theory in which the geometry of an additional dimension is fractal has been considered. In such a theory the mass of an elementary electric charge appears to be many orders of magnitude smaller than the Planck mass, and the "tower" of masses which correspond to higher integer charges becomes aperiodic. 


\end{abstract}

]
\narrowtext

\tightenlines

In modern Kaluza-Klein theories \cite{1} the extra N-4 space-time dimensions are considered to be compact and small (with characteristic size on the order of the Planck length). The symmetries of this internal space are chosen to be the gauge symmetries of some gauge theory \cite{2}, so a unified theory would contain gravity together with the other observed fields. In the original form of the theory a five-dimensional space-time was introduced where the four dimensions $x^1, ..., x^4$ were identified with the observed space-time. The associated 10 components of the metric tensor $g_{\alpha \beta }$ were used to describe gravity. After a compactified fifth dimension $x^5$ with a small circumference $L$ was added, the extra four metric components $g_{\alpha 5}$ connecting $x^5$ to $x^1, ..., x^4$ gave four extra degrees of freedom which were interpreted as the electromagnetic potential (here and everywhere else in the paper we use the following convention for greek and latin indices: $\alpha = 1, ..., 4$; $i = 1, ..., 5$). An additional scalar field $g_{55}$ or dilaton may be either set to a constant, or allowed to vary. 

When a quantum field $\psi $ coupled to this metric via an equation

\begin{equation} 
\Box_5 \psi +a\psi =0
\end{equation}

is considered, where $\Box_5$ is the covariant five-dimensional d'Alembert operator, the solutions for the field $\psi $ must be periodic in the $x^5$ coordinate. This leads to the appearance of an infinite "tower" of solutions with quantized $x^5$-component of the momentum: 

\begin{equation}
q^5_n=2\pi n/L 
\end{equation}

where n is an integer. In our four-dimensional space-time on a large scale such solutions with $n\neq 0$ interact with the electromagnetic potential $g_{\alpha 5}$ as charged particles with an electric charge $e_n$ and mass $m_n$:

\begin{equation}
e_n = \hbar q_n(16\pi G)^{1/2}/c
\end{equation}

\begin{equation}
m_n = \hbar (q_n^2-a)^{1/2}/c
\end{equation}

where $G$ is the gravitational constant (see for example the derivation in \cite{3}). In this theory the conservation of the electric charge is a simple consequence of the conservation of the $x^5$-component of the momentum. Although a unified description of gravity and electromagnetism has been achieved, the result obtained for the mass of an elementary electric charge is unsatisfactory large if $L$ is of the order of the Planck length, and the constant $a$ is small. Fine tuning of $a$ to 20 decimal places may bring the mass to a reasonably low value, but such an arbitrary fine tuning is hardly a satisfactory solution to the problem. 

Similar problems remain in all the modern versions of Kaluza-Klein theory, which introduce up to a dozen additional space-time dimensions and sometimes assume a complicated topology of the "internal space". Despite this theoretical diversity, a common feature of all the Kaluza-Klein theories introduced so far is the integer number of additional dimensions used. But what if the topological properties of the "internal space" are substantially different from the properties of our familiar four-dimensional space-time on a large scale? After all, we make a huge leap of faith in assuming that space-time on the Planck scale remains "smooth" and differentiable. It is reasonable at least to consider the possibility that on the Planck scale space-time experiences such substantial quantum fluctuations that it may be better described by fractal geometry with some non-integer dimension (some attempts to consider fractal space-time geometries may be found in the literature, see for example \cite{4} and the references therein). What would be the consequences of the fractal extra dimensions in the Kaluza-Klein theory? The goal of this paper is to answer this question qualitatively in the most simple situation of a fractal extra dimension $D$ within the range $0<D<2$.    

Fractal objects and fractal dimensions are very useful mathematical concepts. A typical problem where the fractal dimension arises naturally is an attempt to measure the perimeter of an island in the ocean. The result would depend on the resolution used in the measurements. The value of the perimeter measured on the large scale from an aerial map would be much smaller than the value obtained by walking along the beach with a ruler, when every tiny curve of the beach is measured. The fractal dimension $D$ is defined from the variation with resolution of the main fractal variable (a length $L$ of a fractal curve, an area of a fractal surface, etc.)\cite{5}. If $D_T$ is the topological dimension ($D_T=1$ for a curve, $D_T=2$ for a surface), the scale dimension $\delta =D-D_T$ is defined as

\begin{equation}
\delta = \frac{d (ln L)}{d (ln(l/\lambda ))},
\end{equation}

where $\lambda $ is the resolution of the measurements. If $\delta $ is constant we obtain

\begin{equation}
L=L_0(l/\lambda )^{\delta },
\end{equation} 

where the length $L_0$ is measured when $\lambda =l$. Let us assume that the circumference of the internal $x^5$ coordinate obeys expression (6) on a sufficiently small resolution scale instead of been constant. As before, the internal space must look small from our four-dimensional macroscopic point of view, so we may assume $L_0$ to be of the order of the Planck length when measured on a scale compatible with the electron's Compton length $l_c=2\pi \hbar /(m_ec)$ or larger (here we are not concerned with the behavior of $\delta $ on macroscopic scales, so we may assume it to be constant or $\delta \rightarrow 0$ on the large scale). An immediate consequence of (6) in the Kaluza-Klein theory is the drastic reduction of mass of an elementary electric charge for $\delta>0$. Qualitatively this is clear from the following simple arguments.

If we start from the circumference of the internal space $L_0$ measured at large scale and small energies, we would obtain the expression (4) for the elementary charge mass which corresponds to $q=2\pi /L_0$. A particle with such a large mass would have the Compton wavelength of the order of the Planck length. It should "see" quite a different circumference of the internal space. For $\delta >0$ the internal space circumference $L$ measured on such a small scale will be much larger than $L_0$. Thus, $qL>>1$ and such a solution can not be the ground state. The elementary electric charge solution must be obtained in a self-consistent manner, and will correspond to a $x^5$-component of the momentum, which is much smaller than $\hbar /L_0$.

In order to find the ground state self-consistently we must find solutions of the equation $q_n L(q_n) = 2\pi n$, which describes field solutions periodic in the $x^5$-coordinate, where $L(q_n)$ satisfies (6). We may assume that the measurement scale $\lambda $ in (6) corresponds to $2\pi /q_n$, since any constant factor may be included in the value of $l$. Thus, we obtain  

\begin{equation}
q_nL_0(\frac{lq_n}{2\pi })^{\delta }=2\pi n
\end{equation} 

Assuming $a=0$ the spectrum of mass looks like

\begin{equation}
m_n=\frac{2\pi \hbar }{c}(\frac{n}{L_0l^{\delta }})^{1/(1+\delta )}
\end{equation}

This "tower" of solutions is quite different from the "tower" (4) obtained in the regular five-dimensional Kaluza-Klein theory. It is no longer periodic at $a=0$, and for $\delta >0$ the elementary charge mass $m_1$ is much smaller than the Planck mass: $m_1<<\frac{2\pi \hbar }{cL_0}$. For example, in the case of $\delta =1$ we obtain 

\begin{equation}
m_1=\frac{2\pi \hbar }{c(L_0l)^{1/2}}.
\end{equation}

Both these developments are good, since they bring the theoretical picture closer to physical reality. 

The qualitative result obtained above finds support in the following more precise theoretical consideration. Since it is difficult to define a length interval along a fractal $x^5$-direction, let us introduce an angular coordinate $\phi ^5$ varying within an interval from 0 to $2\pi $. Despite the fractal character of $x^5$ we may unambiguously define $\phi ^5$ for each point $A$ as 

\begin{equation}
\phi ^5 = 2\pi \frac{x^5_0}{L_0},   
\end{equation}

where $x^5_0$ is the distance along the $x^5$-coordinate from point $A$ to some point $A_0$ (designated as a zero-point) measured on a large scale, and $L_0$ is the circumference along the $x^5$-direction measured on a large scale. Although all the coordinates are supposed to be treated equally on the small scale, we leave the question of the possible fractal nature of the other four coordinates of our common space-time out of consideration, and write the metric as

\begin{equation}
ds^2=g_{\alpha \beta }dx^{\alpha }dx^{\beta }+2g_{\alpha 5}dx^{\alpha }d\phi ^5+g_{55}d\phi ^5d\phi ^5,
\end{equation} 

where $g_{\alpha 5}$ depend explicitly on $x^{\alpha }$ and the scale of measurements, and $g_{55}$ also depends explicitly on the measurements scale.
This dependence on the scale accounts for the fractal nature of the $x^5$-coordinate. Here we are not interested in possible spatial dependence of $g_{55}$ and consider all $g_{i5}$ components to be independent of $\phi ^5$. Scale dependence of the metric will be addressed later in the discussion. In any event, we consider all $g_{i5}$ to be small at all scales, so $ds^2$ remains reasonably well defined in our four-dimensional space-time.

An analog of equation (1) with $a=0$ for the quantum field $\psi $ in this metric is

\begin{eqnarray}
\frac{\partial }{\partial x^{\alpha }}(g^{\alpha \beta }\frac{\partial \psi }{\partial x^{\beta }})+\frac{\partial }{\partial x^{\alpha }}(g^{\alpha 5}\frac{\partial \psi }{\partial \phi ^5})+ \nonumber \\
\frac{\partial }{\partial \phi ^5}(g^{5\alpha }\frac{\partial \psi }{\partial x^{\alpha }})+\frac{\partial }{\partial \phi ^5}(g^{55}\frac{\partial \psi }{\partial \phi ^5})=0
\end{eqnarray}

Since we assume that the $g_{i5}$ do not depend explicitly on $\phi ^5$, we should not address the meaning of terms like $\partial g^{i5}/\partial \phi ^5$, which would be ambiguous for a fractal $x^5$-coordinate. Thus, equation (12) remains well defined, and we may search for its solution in the usual form as
$\psi = \Psi (x^{\alpha })e^{iq\phi ^5}$, where periodicity in $\phi ^5$ requires $q_n=n$. As a result, we obtain

\begin{equation}
\Box \psi -q_n^2\frac{1-g_{\alpha 5}g^{\alpha 5}}{g_{55}}\psi+2iq_ng^{\alpha 5}\frac{\partial \psi}{\partial x^{\alpha }}+iq_n \frac{\partial g^{\alpha 5}}{\partial x^{\alpha }}\psi = 0
\end{equation} 

This is the same as the Klein-Gordon equation in the presence of an electromagnetic field: in four-dimensional space-time it describes a particle of mass

\begin{equation}
m=\frac{2\pi \hbar q_n}{cg_{55}^{1/2}(q_n)}
\end{equation}

which interacts with a vector field $g^{\alpha 5}$ through a quantized charge $e_n \sim q_n$. If we identify $g_{55}^{1/2}(q)$ as the scale dependent circumference $L(q)$ of the internal space, equations (14) and (8) will be equivalent to each other, and we arrive at the same result for the elementary charge mass and the aperiodic "tower" of solutions described qualitatively earlier. 

Thus far the question of scale dependence of $g_{i5}$ components in (11) has been left without detailed consideration. In order for $ds^2$ to remain a four-dimensional scalar these components must be functions of other four-dimensional scalars. Thus, a natural choice of the scalar scale would be the Compton wavelength $l_c=2\pi \hbar /(mc)$, where $m$ is the effective four-dimensional mass, obtained similar to (14). This choice is consistent with the choice of scale that led to (7). Thus, despite the limited and heuristic nature of our approach (we left the question of possible fractal nature of the other four space-time coordinates without any consideration) some self-consistency has been achieved.

The numerical value of mass of an elementary charge obtained from (8) or (9) remains quite large. If we assume $l$ to be equal to the electron's Compton length equation (9) gives $m_1\sim 10^{11}m_e$. This estimate may be reduced by a few orders of magnitude by selecting somewhat larger values of $L_0$ and $l$, or by increasing the scale dimension $\delta $, so the effect of fractal extra dimension may in principle show up at an energy scale of hundreds of TeV. Overall, addition of the fractal extra dimension appears to be an alternative way of introducing large extra dimensions which has become very popular recently \cite{6}.

In conclusion, we have considered a Kaluza-Klein theory where the geometry of an extra dimension is fractal. In such a theory the mass of an elementary electric charge appears to be many orders of magnitude smaller than the Planck mass, and the "tower" of masses which correspond to higher integer charges becomes aperiodic.


\begin{references}

\bibitem{1} T. Kaluza, Preus. Acad. Wiss. K 1, 966 (1921); O. Klein, Z.Phys. 37, 895 (1926).

\bibitem{2} M.J. Duff, B.E.W. Nilsson, and C.N. Pope, Phys.Rep. 130, 1 (1986).

\bibitem{3} A. Chodos and S. Detweiler, Phys.Rev. D 21, 2167 (1980).

\bibitem{4} L.Y. Kobelev, physics/0006029.

\bibitem{5} B. Mandelbrot, Fractals: form, chance, and dimension (Freeman, San Francisco, 1977).

\bibitem{6} L. Randall and R. Sundrum, Phys.Rev.Lett. 83, 3370 (1999).

\end{references}
\end{document}